\documentclass[12pt,a4paper]{article}
\usepackage{amsmath}
\usepackage[dvips]{graphicx}
\begin{document}

\title{Direct measurement of intersection  angle of invariant manifolds
for area preserving mappings}
\author{Tetsuro KONISHI \thanks{\tt tkonishi@allegro.phys.nagoya-u.ac.jp} \\
Department of Physics, School of Science,\\
Nagoya University, 464-8602, Nagoya, Japan}

\maketitle

\begin{abstract}
  Intersection angles of stable and unstable manifolds for area preserving 
  mappings are numerically calculated by extremely accurate computation.
  With the use of multiprecision library the values of angle as small as
  $10^{-400}$ are obtained.
  The singular   dependence of the angle on the magnitude of 
  hyperbolicity is confirmed.
The power-law type prefactor with Stokes constant is also in good agreement
with analytical estimation.
\end{abstract}
\section{Introduction}

Invariant manifolds are essential components in 
the phase space structure of Hamiltonian systems. A stable  manifold
$W^s$  and an unstable manifold $W^u$ of
a fixed point (or a periodic point)
is a set of points which asymptotically go to the fixed point as
$t\rightarrow\infty$ and $t\rightarrow -\infty$, respectively.
These manifolds
are invariant with time and are called invariant manifolds. 

Suppose we have an integrable Hamiltonian
system $H_0(I)$ 
where an unstable manifold is smoothly connected to a stable manifold
to form a separatrix.
 If we perturb the system as 
$H = H_0(I) + \varepsilon H_1(I,\varphi)$ then the system is 
no longer integrable in general and the unstable manifold and stable
manifold do not
connect smoothly anymore. Rather they intersect transversally
and cross each other infinitely many times to produce a complex structure
called homoclinic tangle, which eventually gives 
rise to chaos~\cite{LL,Tabor}.
 Hence the intersection of invariant manifolds are
quite important as the origin of Hamiltonian chaos.

\begin{figure}[htbp]
  \includegraphics[width=8cm]{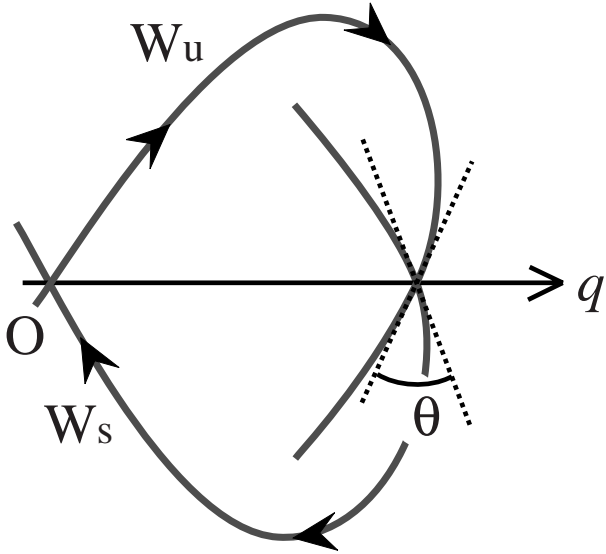}
  \caption{Intersection of unstable and stable manifolds  and intersection 
angle $\theta$. The origin O is a hyperbolic fixed point.}
  \label{fig:intersection}
\end{figure}
At the point of intersection between $W^u$ and $W^s$ we can define the
``intersection angle'' (which is also called as a splitting angle) 
 \, $\theta$ \, between these two manifolds. For several
important cases it is estimated to be quite small and behave singularly 
with respect to the maginitude of hyperbolicity  $\varepsilon$ 
of the fixed 
point~\cite{lazutkin-89}.
Its leading order behavior for $0 < \varepsilon \ll 1$ is  
\begin{equation}
  \label{eq:singular}
  \theta = c_1 \varepsilon^{-\mu}\exp\left(-\frac{c_2}{\varepsilon}\right) \ ,
\end{equation}
where $c_1$, $c_2 (>0 )$ and $\mu (>0) $  are constants.
For small values of $\varepsilon$ the exponential factor makes the
angle $\theta$ extremely small.
Recent  asymptotic calculation of functional form of invariant manifolds
also contain this singular factor~\cite{hakim-mallick-93,tovbis-prep,
tovbis-chaos,nakamura-hamada-96,hirata-4d}
thus supporting the   estimation.

For area preserving mappings, the area enclosed by an 
unstable and a stable manifold is equal to the flux from inside to 
outside (or from outside to inside) of the separatrix per one time step.
If the intersection angle gets small so does the area. Hence the smallness
of the intersection angle is also important in understanding the 
global dynamics and relaxation process in Hamiltonian systems.

The singularly smallness of the intersection angle is, however, 
not yet confirmed thoroughly. 
Analytical estimation mentioned above is asymptotic
approximation, and since it is  asymptotic expression for
$\varepsilon  \rightarrow 0$, we do not know from what value of 
$\varepsilon$ we can observe the singular behavior.

Usually numerical computation can support analytical estimation.
However, ordinary numerical computation is useless in this case
to check the singular behavior because of limited precision.
Suppose we would like to confirm the proportionality of 
intersection angle as $\exp(-\pi^2/\varepsilon)$. If $\varepsilon=0.1$
the factor is about $\exp(-100)\sim 10^{-43}$.
Let us write the normalized tangent vector and their components
 for stable and unstable manifolds as
$\vec{t}^s = (\alpha, \beta)$ and $\vec{t}^u = (\alpha + d\alpha,
\beta + d\beta)$ respectively, and the intersection angle as $\theta$.
Then $|d\alpha| = |\beta\theta| $ and $|d\beta|=|\alpha\theta|$ up to 
the first order of $\theta$. Hence if the angle is of the order of 
$10^{-40}$ then  the  differences between  
values of the components of 
two vectors $\vec{t}^s$ and $\vec{t}^u$ are  far beyond the 
range of ordinary double precision of 64-bit, where mantissa
(fractional part) is at most 16 decimal digits.

There are two methods which we may use for  problems which require 
extremely high precision. One is ``validated
 computation''~\cite{validated} and 
the other is ``multiprecision''. ``Validated computation'' is a 
method to obtain {\it rigorous} results with numerical computation.
It is based on operation for intervals rather than for numbers. 
Usual numerical computation on real numbers
says e.g., $x=0.123$, but this just mean that
the value of $x$ will be  near 0.123 and it does not mean that the value
of $x$ is exactly 0.123.
On the other hand,  with the use of validated computation 
we can rigorously say that the value of $x$ satisfies , 
for example,  $ 0.122 <  x <   0.124$. 

Although  validated computation is a powerful tool, it seems not to be
all-purpose,
 and it seems to  take time to reformulate the problem one want to 
solve to fit validated computation. Hence we adopt the second tool,
multiprecision library.

In usual numerical computation the precision of 
variables are uniquely defined in  each implementation of 
 language ({\tt C, FORTRAN}, etc.).
On the other hand,
using multiprecision library users can define the precision of 
variables arbitrarily. For various multiprecision tools see
the {\tt hfloat} web page~\cite{hfloat}.
Here we adopt a multiprecision library called {\tt cln}~\cite{cln}
which can be used in {\tt C++}.

The purpose of this paper is (i) to numerically confirm the singular
 dependence of the intersection angle using multiprecision library,
(ii) to find the range of $\varepsilon$ where the singular dependence
appears. 

There have been several attempts to numerically compute the singular
behavior of the intersection angle. In the paper ~\cite{lazutkin-89}
Lazutkin et al. cites a numerical result by C. Sim\'{o} as private 
communication and writes that the value of the angle 
by their analytical estimation is
 ``in good accordance with numerical data''.
Also in \cite{kadanoff-beyondallorder-92}, Amick et al.
computed the splitting distance of manifolds for
a 
4th order  map 
$x_{t+3} = x_t + \varepsilon g(x_{t+2})$, $g(x)\equiv x(1-x) $ with 
ordinary double precision.
In this paper, 
we give a refined  and concrete result for these pioneering work,
by performing direct computation with multiprecision,

This paper is organised as follows. In section 2 we introduce 
symplectic mappings 
for which we calculate the intersection angles. In section 3 
we describe the method of computation, and the result is presented 
in section 4. In section 5 we give summary and discussions.

\section{Models}

The first model we use is a ``double-well map'';
\begin{equation}
  \phi_{\rm dw} \,:\, (q,p)\mapsto (q',p') \ ,  \ 
  p' = p - \varepsilon(2q^3 - q) \ ,  \ 
 q' = q + \varepsilon p' \ .            \label{eq:dw}
\end{equation}
This map has a hyperbolic fixed point at (0,0).

For this mapping the intersection angle is estimated to 
be~\cite{nakamura-hamada-96}
\begin{equation}
  \label{eq:angle-double-well}
\theta_{\rm dw} =  \frac{c_{\rm dw}}{\varepsilon^5}
\exp\left(-\frac{\pi^2}{\varepsilon}\right) \  , \ 
c_{\rm dw} = 5.00\times 10^4 \ ,
\end{equation}
after change of coordinate\footnote{The paper 
\protect\cite{nakamura-hamada-96} adopts different notation for the map
(\protect\ref{eq:dw}) hence the value of the angle $\theta$
and the prefactor $c_{\rm dw}$ is  rescaled appropreately in 
eq. (\ref{eq:angle-double-well}).}.

The second model is the well-known standard map~\cite{SM1};
\begin{equation}
  \label{eq:st}
  \phi_{\rm st}\, : \,  (q,p) \mapsto (q',p') \ , \ 
  p' = p + \frac{K}{2\pi}\sin(2\pi q) \ ,  \ 
  q' = q + p' \ .
\end{equation}
This map has hyperbolic fixed points at $(n,0)$ ($n$ : integer). 
For standard map the intersection angle is estimated to be
~\cite{lazutkin-89} 
\begin{equation}
  \label{eq:angle-st}
  \theta_{\rm st} =  \frac{c_{\rm st}}{K}\exp\left(-\frac{\pi^2}{\sqrt{K}}  
\right) \ , \ c_{\rm st} = 1118.827706...  \ .
\end{equation}

\section{Method of calculation}
Here we describe the method of  calculation of intersection angle 
for the double well map (\ref{eq:dw}). The method for standard map
is similar.

In short, we choose two approximate points $P_\pm$ for an intersection point, 
and approximate the intersection angle 
 by the ones between the two manifolds which are on $P_\pm$.

\subsection{finding an interval which contains PIP} 

First we search for an intersection point of $W^u$ and $W^s$. 
Following an argument similar to \cite{gelfreich-st-94}, it is easily shown that
one of the intersection points called principal intersection point (PIP)
is on the $q$-axis.
\begin{figure}[hbtp]
  \begin{center}
    \includegraphics[width=12cm]{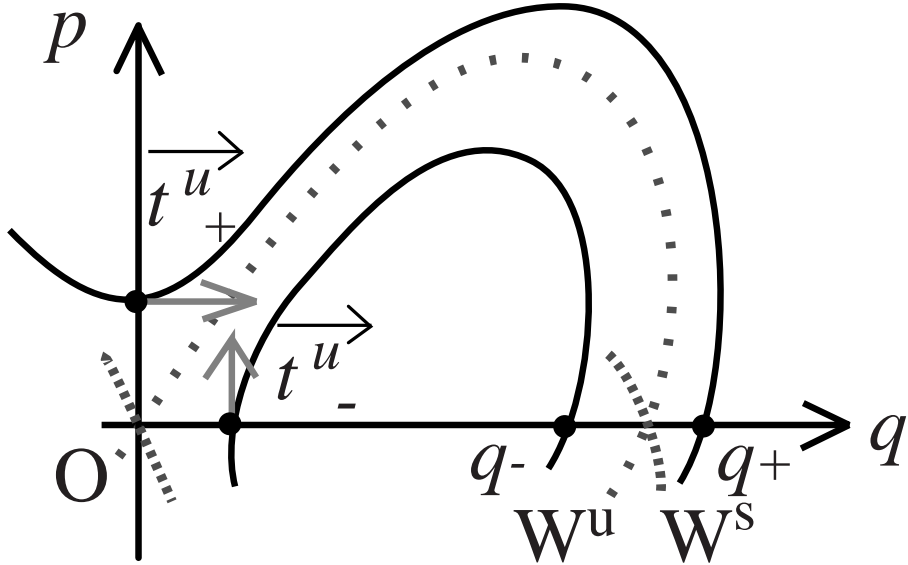}
    \caption{True stable and unstable manifolds (dashed lines) and 
approximate manifolds (solid lines) and tangent vectors}
    \label{fig:tangent-vectors}
  \end{center}
\end{figure}

Suppose we choose an initial point on $q$-axis and update the point
backward in time. If the point is near but not on the stable manifold 
it will approach to the hyperbolic fixed point and then goes away from it.
The direction the point goes away shows whether 
the initial point is on the right or left of the true intersection point.
Hence we can set an interval on $q$-axis which include the
intersection point between the stable and unstable manifolds.

Then  we make the interval narrower by bisection. 
We call the final interval as $(q_{-}, q_{+})$. 
In actual calculation 
we set the final width of the interval as 
\begin{equation}
  \label{eq:q-width}
  |q_{+} - q_{-}| = 10^{-100} \ \  \text{for} \  \varepsilon \ge 0.02 
\end{equation}
and
\begin{equation}
  |q_{+} - q_{-}| = 10^{-800} \ \ \text{for} \  \varepsilon = 0.01 
\end{equation}
We will use the points $(q_{-},0)$ and $(q_{+},0)$ as approximations for
the intersection point and denote as $P_{-}$ and $  P_{+}$
respectively, i.e.,
\begin{equation}
  \label{eq:pip-approx}
  P_\pm \equiv (q_\pm,0) 
\end{equation}

\subsection{construction of tangent vectors in the linear region} 
 Now we are going
to make  tangent vectors at these approximate points $P_\pm$.

Although we do not know directly the tangent vectors at 
$P_\pm$, we can calculate them from the linear neighbourhood 
around the hyperbolic fixed point $(0,0)$.

First we update the points $P_\pm$ forward or backward in time;
\begin{equation}
  P^u_\pm \equiv \phi_{\rm dw}^{-N_u}\, \left(P_\pm\right), \ \ 
  P^s_\pm \equiv \phi_{\rm dw}^{N_s} \,\left(P_\pm\right) \ 
\end{equation}

where the time steps $N_u$ and $N_s$ are sufficiently large
for $|P^u_\pm| \ll 1 $ and  $|P^s_\pm| \ll 1 $. In actual 
calculation we determined $N_u$ and $N_s$  so that
$P^u_\pm$ and $P^s_\pm$  be 
the points which are nearest to the hyperbolic fixed point (0,0)
on the orbits $ \phi_{\rm dw}^{-n} \left(P_\pm\right)$
 and $\phi_{\rm dw}^n \left(P_\pm\right)$, 
respectively.

Since  the points $P^u_\pm$ and $P^s_\pm$ are quite near to the  
fixed point (0,0)
the map $\phi_{\rm dw}$ can be approximated by its linearized map;
\begin{equation}
  \label{eq:dw-linearize}
  \phi_{\rm dw}(q,p) = D\phi_{\rm dw}(0,0)
  \begin{pmatrix}
    q \\ p
  \end{pmatrix}
+ O(q^2,p^2) \ .
\end{equation}
Hence the tangent vector of the manifolds which passes $P^u$ or $P^s$ 
can be approximated  by the invariant curve of the linearized map
$ D\phi(0,0)$, that is, hyperbola.
If we denote
\begin{equation}
D\phi_{\rm dw}
= 
\begin{pmatrix}
a+d & c+b \\ c-b & a-d
\end{pmatrix}
\end{equation}
Then the curve
\begin{equation}
  \Phi(q,p) \equiv (b-c)q^2 + 2dqp + (b+c)p^2 = \text{const.}
\label{eq:hyperbola}
\end{equation}
is invariant under $\phi_{\rm dw}$ ~\cite{greene-linearized}.
Hence we can obtain the tangent vector $\overrightarrow{t^u}$ and
$\overrightarrow{t^s}$
 of unstable and stable manifolds as the tangent vector of 
the hyperbola (\ref{eq:hyperbola})
\begin{equation}
  \label{eq:tangent}
  \overrightarrow{t^\gamma}_\pm \equiv \vec{t}(P^\gamma_\pm) = 
  {}^t\left(-\frac{\partial \Phi}{\partial p},\,
    \frac{\partial \Phi}{\partial q}\right)(P^\gamma_\pm) \ , 
\gamma = u \ \mbox{or} \ s \ .
\end{equation}

\subsection{calculation of intersection angles}
The tangent vectors at approximate intersection point $P_\pm$ is 
obtained by pulling back the tangent vectors $\vec{t^u}$, $\vec{t^s}$
to the points $P_\pm$
as
\begin{equation}
  \overrightarrow{v^u}_\pm = D\phi_{\rm}^{N_u} \left(\overrightarrow{t^u}_\pm
  \right)\ , \
  \overrightarrow{v^s}_\pm = D\phi_{\rm}^{-N_s} \left(\overrightarrow{t^s}_\pm\right) \ . \
\end{equation}

Approximation values for 
intersection angle $\theta$ at each approximate intersection point
$P_\pm$ is obtained as
\begin{equation}
  \cos\theta_\pm = \frac{\overrightarrow{v^u}_\pm\cdot\overrightarrow{v^s}_\pm}
  {\left|\overrightarrow{v^u}_\pm\right| \left|\overrightarrow{v^s}_\pm
\right|} \ .
\end{equation}

\section{Results}

Now we present the results. Computation is performed on a
Pentium-II processor
PC with Linux operating system (Vine Linux 1.0b)
 with GNU C/C++ compiler {\tt g++} 2.7.2.3
and {\tt egcs} 1.0.3-release,
and the version of multiprecision library is {\tt cln} 1.0.1.
The computation time is typically several hundred times longer than 
ordinary numerical computation with double precision.

\begin{figure}[hbtp]
  \begin{center}
    \includegraphics[width=10cm]{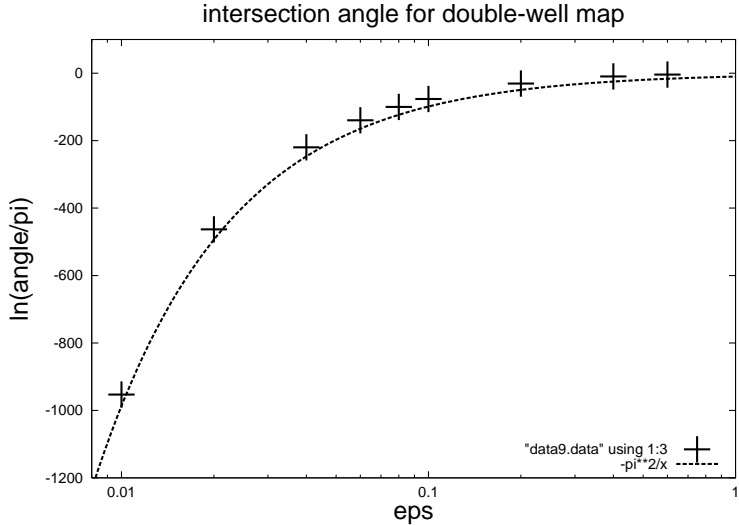}
    \caption{Intersection angle for double well map~(\protect\ref{eq:dw}). 
      Values are in  Table\protect\ref{tab:angle-double-well}.
      The dotted line 
      represents analytical estimation for the singular behavior
      $\exp(-\pi^2/\varepsilon)$(eq.(\protect\ref{eq:angle-double-well})).}.
    \label{fig:angle-dw}
  \end{center}
\end{figure}

The result for double well map~(\ref{eq:dw}) is summarised in the 
Table \ref{tab:angle-double-well}
and Fig.\ref{fig:angle-dw}. Calculation is performed with 1000 decimal digits
for $\varepsilon \ge 0.02$ and 4000 digits for $\varepsilon=0.01$.
\begin{table}[hbtp]
  \begin{center}
    \begin{tabular}{l|ll}
\hline
$\varepsilon$ & $\theta_{-}/\pi$  &  $\theta_{+}/\pi$ \\ \hline
0.6  &   0.00620123      &       0.00620123      \\
0.4  &   2.03305e-05     &       2.03305e-05     \\
0.2  &   1.61224e-14     &       1.61224e-14     \\
0.1  &   2.0686e-34      &       2.0686e-34      \\
0.08 &   1.23026e-44     &       1.23026e-44     \\
0.06 &  7.24697e-62     &       7.24697e-62\\
0.04 &   1.06168e-96     &       1.06168e-96     \\
0.02 &   2.58525e-202    &       2.38495e-202    \\
0.01 &     1.16442e-414  &  1.16442e-414         \\ \hline
    \end{tabular}
    \caption{Intersection angle $\theta$ for double well map.
1000 decimal digits for $\varepsilon \ge 0.02$ and 4000 decimal digits 
for $\varepsilon=0.01$. $|q_L - q_R| = 10^{-100}$ for $\varepsilon\ge 0.02$
and $10^{-800}$ for $\varepsilon=0.01$  . }
    \label{tab:angle-double-well}
  \end{center}
\end{table}

Here we can see that our numerical results agree quite well with the
estimation (eq.(\ref{eq:angle-double-well})). It is clear that the data
cannot be fitted by power law.
Also we note that the singular
have behavior appears for the value of $\varepsilon$ as large as
$\varepsilon=0.1$. 

The difference of the values between $\theta_{-}$ and $\theta_{+}$ 
is an indicator of error. Table~\ref{tab:angle-double-well} shows
the error is small.

Next we see the results for standard map~(\ref{eq:st}).
For standard map we have the principal intersection point on
the line $q=\pi$~\cite{gelfreich-st-94}, hence we take its
 approximation points as $P_\pm = (\pi,p_\pm)$.
Calculation is performed with 1000 decimal digits.
For standard map in the notation (\ref{eq:st})
 the parameter $\varepsilon$ is equal to $\sqrt{K}$.

\begin{table}[hbtp]
  \begin{center}
    \begin{tabular}{lll}
\hline
$K$ &   $\theta_{-}/\pi$  &  $\theta_{+}/\pi$ \\ \hline
1.5    & 0.0692401 &            0.0692401   \\
1.0    & 0.0250374 &            0.0250374   \\
0.8    & 0.0114877 &            0.0114877   \\
0.4    & 0.000324148 &             0.000324148 \\
0.2    & 1.16407e-06 &            1.16407e-06\\
0.1    & 2.6934e-10  &            2.6934e-10   \\
0.08   & 8.60911e-12 &            8.60911e-12\\
0.04   & 9.47743e-18 &            9.47743e-18\\
0.02   & 2.58424e-26 &            2.58424e-26\\
0.01   & 1.46928e-38 &            1.46928e-38\\
0.008  & 1.60962e-43 &            1.60962e-43\\
0.004  & 4.59792e-63 &            4.59792e-63\\
0.002  & 7.84703e-91  &            7.84703e-91\\
0.001  & 3.14686e-130 &           3.14686e-130\\ \hline
    \end{tabular}
    \caption{Intersection angle for standard map.
1000 decimal digits are used. $|p_{+} - p_{-}| = 10^{-100}$. }
    \label{tab:st}
  \end{center}
\end{table}

\begin{figure}[hbtp]
  \begin{center}
    \includegraphics[width=10cm]{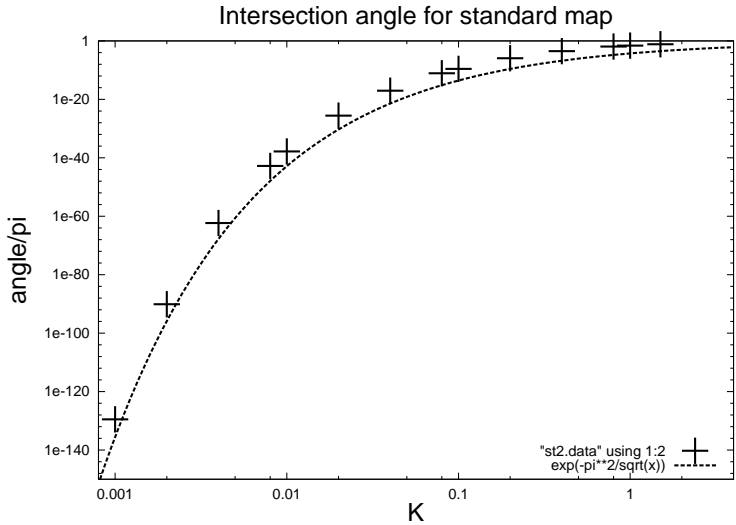}
    \caption{Intersection angle for standard map.
      Values are in  Table~\ref{tab:st}.
      The dotted line 
      represents analytical estimation for the singular behavior
      $\exp(-\pi^2/\sqrt{K})$ (eq.(\protect\ref{eq:angle-st})).}
    \label{fig:angle-st}
  \end{center}
\end{figure}

From the  figure~\ref{fig:angle-st} again we see that the intersection angles 
agree quite well with analytical 
estimation~(\ref{eq:angle-st}).
Note that we apparently have  agreement even above the $K_c$, 
where the last KAM torus
is broken and the phase space has global chaotic sea.

Finally we show the prefactors of the singular behavior
($\displaystyle c_1 \varepsilon^{-\mu}$ in eq.(\ref{eq:singular}).)
Here we examine whether  the prefactor is in fact a  power law type 
with previously estimated exponent $\mu$. Also the value of the constant 
$c_1$ is interesting to check, because $c_1$ contains 
a constant called Stokes constant, which represents the change in the form of
asymptotic expansion of the manifolds\cite{hakim-mallick-93,tovbis-prep,
tovbis-chaos,hirata-4d,nakamura-hamada-96}.

\begin{figure}[hbtp]
  \begin{center}
    \includegraphics[width=9cm]{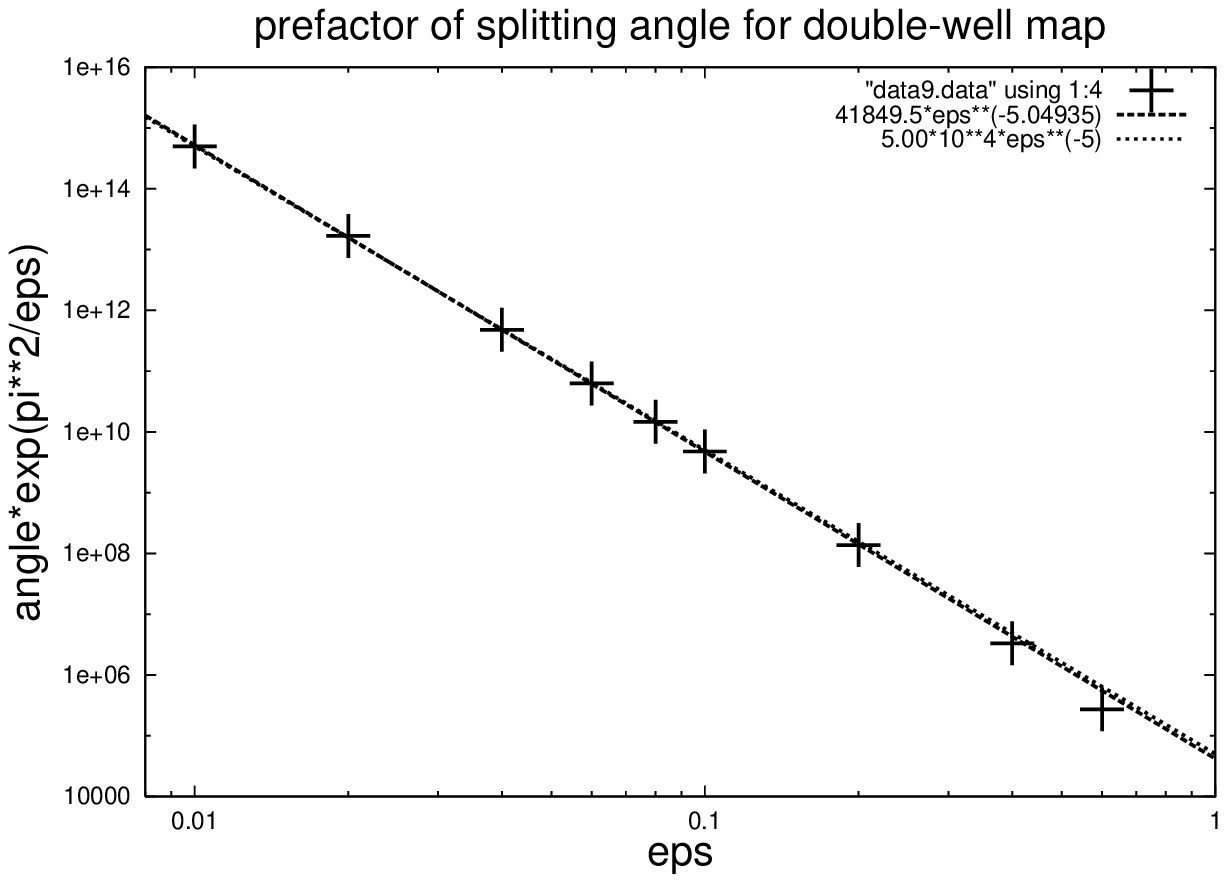}
    \caption{Prefactor of intersection angle for double-well 
      map~(\protect\ref{eq:dw}).
      Dashed line   represents a least square fitting
      of the data for $0.01 \le \varepsilon \le 0.1$, 
      and dotted line represents an analytical
      estimation $\displaystyle c_{\rm dw} /\varepsilon^5$~(\protect\ref{eq:angle-double-well}). } 
    \label{fig:dw-pref}
  \end{center}
\end{figure}

\begin{figure}[hbtp]
  \begin{center}
    \includegraphics[width=9cm]{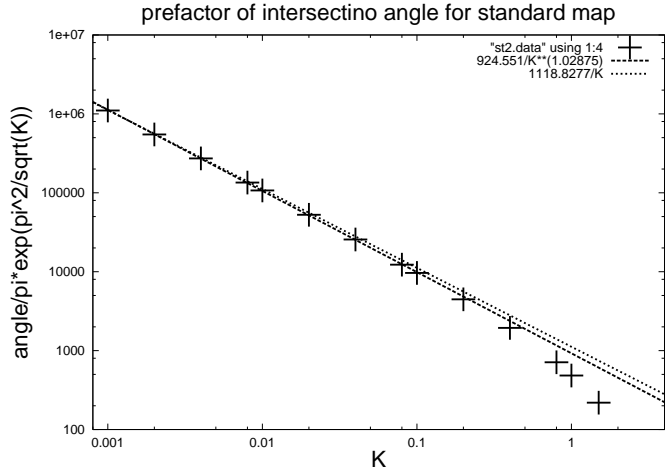}
    \caption{Prefactor of intersection angle of standard map.
      Dashed line   represents a least square fitting
      of the data for $0.001 \le K \le 0.1$ ,
      and dotted line represents an analytical
      estimation $\displaystyle c_{\rm st}/K$ ~(\protect\ref{eq:angle-st}).   }
    \label{fig:st-pref}
  \end{center}
\end{figure}

Fig. \ref{fig:dw-pref}  shows the $\varepsilon$ dependence of the value
$\displaystyle \theta\exp(\pi^2/\varepsilon)$. Here we see a clear 
power-law dependence for small values of $\varepsilon$.
 A least square fitting  for $ 0.01 \le \varepsilon \le 0.1$ gives
 \begin{equation}
   \theta\cdot \exp(\pi^2/\varepsilon) = 41894.5/\varepsilon^{5.04935} \ , 
 \end{equation}
which is in good agreement with the estimation (\ref{eq:angle-double-well}).
The difference in the exponent ( 5.04935 for our numerical result and
5  for \cite{nakamura-hamada-96} ) is negligible.
Also the prefactor 41894.5 is in good agreement with the estimated value 
of $5.00\times 10^4$.

Fig. \ref{fig:st-pref}  is for the standard map, where we see
a fitting for $ 0.001\le K \le 0.1$
\begin{equation}
  \label{eq:st-pref}
  \theta/\pi\exp(\pi^2/\sqrt{K}) = 924.551/K^{1.02875} \ ,
\end{equation}
again in good agreement with the estimation (\ref{eq:angle-st}). 
The difference between our numerical result and the estimation
is negligible, as shown in
the figure. 
\section{Summary and discussion}
In this paper we have numerically computed the intersection angles between
stable and unstable manifolds for 
two area preserving mappings.
Computation is performed with extremely high precision 
( with 1000 or 4000 decimal digits)
 with the use of multiprecision library {\tt cln}
to obtain the angle as small as $10^{-400}$. 
The singular behavior
previously obtained by analytical estimation is confirmed.
Hence we have confirmed that the singular
small behavior of the intersection angle $\theta \propto \exp(-c/\varepsilon)$
indeed exists for area preserving mappings. Thus the flux in the phase space 
is quite small for small values of $\varepsilon$, which will make the relaxation
process quite slow.
In addition, although the analytical
 estimation is obtained as the asymptotic approximation
formula for $\varepsilon \rightarrow 0$ , our numerical results show that
the singular dependence appears from rather large value of $\varepsilon$.

Also we have measured the prefactor of the singular factor, and confirmed
that the analytical estimations for  prefactors are correct.
Exponent of the power law is in almost perfect agreement.
It appears constants in the   prefactors slightly differ from the
ones in  analytical estimation.
This does not necessarily imply  that the estimation need to be 
refined.
For  the analytical estimation contains a free parameter 
called switching function  $S(0)$ 
which take a value between 0 and 1\cite{nakamura-hamada-96}. 
Although the estimation is given for $S(0)=1/2$, $S(0)$ can be 
some other value. In fact \cite{nakamura-hamada-96} takes the value
$S(0)=0.3$ when comparing their functional form of unstable manifold
with numerically obtained orbit.
Note that within the range of parameters computed the constant factor
do not appear to depend on $\varepsilon$ or $K$. 

The singularly smallness of  intersection angle was
one of several things in the study of Hamiltonian chaos
which are predicted but not yet confirmed precisely. Among others are 
Nekhoroshev bound~\cite{Nekhoroshev,lochak} and Arnold diffusion~\cite{Arnold}.
In both cases singular factor appears and it is believed that the singular
behavior accounts for slow motion in high-dimensional
nearly integrable Hamiltonian systems~\cite{nekh-size}.  
Extreme precision computation using
multiprecision library will  reveal some new properties about the slow dynamics of 
high-dimensional Hamiltonian systems.

\section{Acknowledgement}
We would like to thank Kazuhiro Nozaki, Yoshihiro Hirata and
the members of R-Lab., Nagoya University 
for stimulating discussions and valuable comments.
Thanks are also due to Bruno Haible, 
the author of the multiprecision libary {\tt cln}.


\end{document}